\documentclass[12pt,preprint]{aastex}
\usepackage{subfigure}
\usepackage{url}

\shorttitle{Imaging and spectral study on the null point of a fan-spine structure}
\shortauthors{Yang et al.}

\begin{document}

\title{Imaging and spectral study on the null point of a fan-spine structure during a solar flare}

\author{Shuhong Yang\altaffilmark{1,2}, Qingmin Zhang\altaffilmark{3,1}, Zhi Xu\altaffilmark{4}, Jun Zhang\altaffilmark{5,1}, Ze Zhong\altaffilmark{6,7} and Yang Guo\altaffilmark{6}}

\altaffiltext{1}{CAS Key Laboratory of Solar Activity, National Astronomical Observatories, Chinese Academy of Sciences, Beijing
100101, People's Republic of China; shuhongyang@nao.cas.cn}

\altaffiltext{2}{School of Astronomy and Space Science, University of Chinese Academy of Sciences, Beijing 100049, People's Republic of China}

\altaffiltext{3}{Key Laboratory of Dark Matter and Space Science, Purple Mountain Observatory, Chinese Academy of Sciences, Nanjing 210033, People's Republic of China}

\altaffiltext{4}{Fuxian Solar Observatory, Yunnan Observatories, Chinese Academy of Sciences, Kunming 650011, People's Republic of China}

\altaffiltext{5}{School of Physics and Materials Science, Anhui University, Hefei 230601, People's Republic of China}

\altaffiltext{6}{School of Astronomy and Space Science and Key Laboratory of Modern Astronomy and Astrophysics in Ministry of Education, Nanjing University, Nanjing 210023, People's Republic of China}

\altaffiltext{7}{Max Planck Institute for Solar System Research, Justus-von-Liebig-Weg 3, D-37077 G{\"o}ttingen, Germany}

\begin{abstract}

Using the multi-instrument observations, we make the first simultaneous imaging and spectral study on the null point of a fan-spine magnetic topology during a solar flare. When magnetic reconnection occurs at the null point, the fan-spine configuration brightens in the (extreme-)ultraviolet channels. In the H$\alpha$ images, the fan-spine structure is partly filled and outlined by the bi-directional material flows ejected from the reconnection site. The extrapolated coronal magnetic field confirms the existence of the fan-spine topology. Before and after the flare peak, the total velocity of the outflows is estimated to be about 60 km s$^{-1}$. During the flare, the Si~{\sc{iv}} line profile at the reconnection region is enhanced both in the blue-wing and red-wing. At the flare peak time, the total velocity of the outflows is found to be 144 km s$^{-1}$. Superposed on the Si~{\sc{iv}} profile, there are several deep absorption lines with the blueshift of several tens of km s$^{-1}$. The reason is inferred to be that the bright reconnection region observed in Si~{\sc{iv}} channel is located under the cooler material appearing as dark features in the H$\alpha$ line. The blueshifted absorption lines indicate the movement of the cooler material toward the observer. The depth of the absorption lines also depends on the amount of cooler material. These results imply that this kind of spectral profiles can be used as a tool to diagnose the properties of cooler material above reconnection site.

\end{abstract}


\section{Introduction}

Magnetic reconnection is a fundamental physical process during which magnetic energy is converted to the kinetic and thermal energy of plasma (Zweibel \& Yamada 2009; Pontin 2011; Yang et al. 2011, 2015a; Yang \& Xiang 2016). Fan-spine magnetic topology is favorable for the occurrence of solar flares through null-point reconnection (Lau \& Finn 1990; Antiochos 1998; Aulanier et al. 2000; Priest \& Titov 1996; Vemareddy \& Wiegelmann 2014; Wyper et al. 2017, 2018; Li et al. 2018; Shen et al. 2019). A fan-spine configuration can be formed through the emergence of a magnetic bipole into the pre-existing unipolar field. Then the minority polarity is surrounded by the magnetic fields with the opposite polarity and thus becomes a parasitic core (Moreno-Insertis et al. 2008; T{\"o}r{\"o}k et al. 2009; Sun et al. 2013; Hou et al. 2019). In the fan-spine topology, a dome-shaped fan divides two different connectivity domains, and the inner spine meets the outer one at the null point (Shibata et al. 1994; Fletcher et al. 2001; Liu et al. 2011; Pontin et al. 2013).

When magnetic reconnection takes place at the null point of a fan-spine, the accelerated particles move along the fan surface and hit the lower atmosphere, where a circular ribbon would be observable in the chromosphere (Masson et al. 2009, 2017; Reid et al. 2012; Zhang et al. 2016, 2020; Zhong et al. 2019; Lee et al. 2020) and even in the photosphere (Hao et al. 2017; Song \& Tian 2018) if the flare is strong enough. When the spine is open to the interplanetary space, a jet and a consequent coronal mass ejection may be formed (Pariat et al. 2009; Hong et al. 2017). Sometimes the spine extends to the remote solar surface, and thus the material expelled from the null point flows toward the far end of the spine, resulting in a remote brightening (Wang \& Liu 2012; Deng et al. 2013; Hernandez-Perez et al. 2017; Li et al. 2017; Yang et al. 2018). As revealed by many studies, the rising filament (or sigmoid) embedded within the fan dome plays an important role in triggering the magnetic reconnection at the null point, and consequently the filament material can be ejected into the large-scale open or distantly closed field lines (e.g., Joshi et al. 2015; Liu et al. 2015; Xu et al. 2017).

Although the fan-spine morphology has been extensively investigated, the previous studies are mainly dedicated to the formation and changes of magnetic configuration and the associated explosive events (e.g., Guglielmino et al. 2010; Jiang et al. 2015; Yang et al. 2015c; Lim et al. 2017; Kumar et al. 2019), and there is no simultaneous imaging and spectral study on the null point during solar flare. In this paper, we study a fan-spine structure in active region (AR) 12736 on 2019 March 22, and focus on the null point where magnetic reconnection occurred during a B6.7 flare, combing the high-quality vector magnetograms from the \emph{Hinode} (Kosugi et al. 2007) with the high-resolution imaging and spectral observations from the New Vacuum Solar Telescope (NVST; Liu et al. 2014), the \emph{Solar Dynamics Observatory} (\emph{SDO}; Pesnell et al. 2012), and the \emph{Interface Region Imaging Spectrograph} (\emph{IRIS}; De Pontieu et al. 2014).

\section{Observations and Data Analysis}

The NVST at the \emph{Fuxian Solar Observatory} of China was pointed to AR 12736 on 22 March 2019. The H$\alpha$ and TiO channels are used to image the highly dynamic activities in the chromosphere and the fine-scale structures in the photosphere, respectively (Yang et al. 2014a,b, 2015b, 2019; Yan et al. 2020). The H$\alpha$ filter is a tunable Lyot filter with a bandwidth of 0.25 {\AA}, which can scan spectra in the $\pm$ 5 {\AA} range with a step size of 0.1 {\AA}. In the present study, we use the H$\alpha$ observations at the line-center (6562.8 {\AA}) and two line wings ($\pm$ 0.4 {\AA}) from 00:57 UT to 04:37 UT. Their pixel size is 0.{\arcsec}136 and cadence is 29 s. In addition, we employ the simultaneous photospheric TiO 7058 {\AA} images with a pixel size of 0.{\arcsec}052 and a cadence of 30 s. The fields of view (FOVs) of the H$\alpha$ and TiO images are 126{\arcsec} $\times$ 126{\arcsec} and 124{\arcsec} $\times$ 100{\arcsec}, respectively. All the data are calibrated from Level 0 to Level 1, including flat field correction and dark current subtraction. Furthermore, the Level 1 data are reconstructed to Level 1+ by speckle masking (Weigelt 1977; Lohmann et al. 1983).

The simultaneous \emph{SDO} observations are also adopted. The Atmospheric Imaging Assembly (AIA; Lemen et al. 2012) on board the \emph{SDO} images the Sun in ten (extreme-)ultraviolet (EUV/UV) channels with a cadence of (12)24 seconds and a spatial sampling of 0{\arcsec}.6 pixel$^{-1}$. The Helioseismic and Magnetic Imager (HMI; Scherrer et al. 2012; Schou et al. 2012) on board the \emph{SDO} provides the full-disk line-of-sight (LOS) magnetograms with a cadence of 45 seconds and a pixel size of 0{\arcsec}.5. They are calibrated to Level 1.5 by using the standard routine under the Solar Software package, and then differentially rotated to the reference time of 02:00 UT.

The \emph{IRIS} observed AR 12736 from 01:43:27 UT to 02:42:16 UT. We use a series of slit-jaw imager (SJI) 1330 {\AA} images with a cadence of 14 s and the simultaneous spectral profiles of Si~{\sc{iv}} 1393.76 {\AA}. The 1330 {\AA} images have a FOV of 60{\arcsec} $\times$ 60{\arcsec} and a plate scale of 0.{\arcsec}166 $\times$ 0.{\arcsec}333. The spectral profiles of Si~{\sc{iv}} are obtained along the slit at the same location with solar-rotation tracking. The pixel size along the slit is 0.{\arcsec}333, and the spectral dispersion is 25.4 m{\AA} pixel$^{-1}$. The absolute wavelength calibration is performed assuming the cold chromospheric S~{\sc{i}} 1401.51 {\AA} on the line profiles in the relatively quiet region to be at rest (Tian et al. 2015, 2018).

The Spectro-Polarimeter (SP; Lites et al. 2013) of the Solar Optical Telescope (Ichimoto et al. 2008; Tsuneta et al. 2008; Suematsu et al. 2008) on board the \emph{Hinode} provides full Stokes profiles of Fe~{\sc{i}} 6301.5 {\AA} and 6302.5 {\AA} lines. The SP scanned AR 12736 along the east-west direction in the fast map mode from 01:02:05 UT to 01:59:30 UT. The scanning step is 0.{\arcsec}30, the pixel size along the slit is 0.{\arcsec}32, and the FOV is 270{\arcsec} $\times$ 164{\arcsec}. The calibrated Stokes profiles (Lites \& Ichimoto 2013) are fitted using the ``MERLIN" inversion code (Lites et al. 2007) based on the assumption of the Milne-Eddington atmosphere model\footnote[1]{\url{https://www2.hao.ucar.edu/csac}}. There are 36 parameters output from the inversion\footnote[2]{\url{https://sot.lmsal.com/data/sot/level2dd/sotsp_level2_description.html}}, including the three components of vector magnetic field (i.e., field strength, inclination, and azimuth), and the filling factor. The field inclination is defined by the angle between the vector magnetic field and the LOS, the field azimuth is defined by the angle between the transverse field and the heliocentric $x$-axis in the anti-clockwise direction, and the filling factor is the magnetic fill fraction. In the vector magnetic field measurements based on the Zeeman effect, there exists a 180$\degr$ ambiguity in determining the field azimuth. To resolve the 180$\degr$ ambiguity, although many different algorithms have been developed, it is still difficult to make a complete removal (Metcalf et al. 2006; Semel \& Skumanich 1998). In the paper of Metcalf et al. (2006), it was pointed out that ``the methods which minimize some measure of the vertical current density in conjunction with minimizing an approximation for the fields' divergence show the most promise." In the present study, we adopt the improved Nonpotential Magnetic Field Calculation (Georgoulis 2005) method to disambiguate the azimuth angles. With the formulae given by Gary \& Hagyard (1990), the disambiguated vector fields in the image plane are transformed to the heliographic components. Furthermore, the geometric mapping of the magnetograms into the heliographic coordinates is performed.

Using the nonlinear force-free field (NLFFF) modeling (Wheatland et al. 2000; Wiegelmann et al. 2012), we reconstruct the coronal structures with the \emph{Hinode}/SP vector magnetogram as the bottom boundary, before which the observed magnetogram is preprocessed to best suit the force-free conditions (Wiegelmann et al. 2006). The extrapolation is performed in the cubic box of 320 $\times$ 200 $\times$ 200 uniformly spaced grid points with $\Delta x= \Delta y=\Delta z=0.\arcsec62$. For the extrapolated magnetic field, the squashing factor \emph{Q} (D{\'e}moulin et al. 1996; Titov et al. 2002) and twist number $\mathcal{T}_w$ (Berger \& Prior 2006) are calculated with the code of Liu et al. (2016).

Additionally, we use the \emph{Geostationary Operational Environmental Satellite} (\emph{GOES}) 1-min cadence data to examine the variation of soft X-ray 1-8 {\AA} flux.

\section{Results}

\begin{figure*}
\centering
\includegraphics
[bb=73 212 518 623,clip,width=\textwidth]{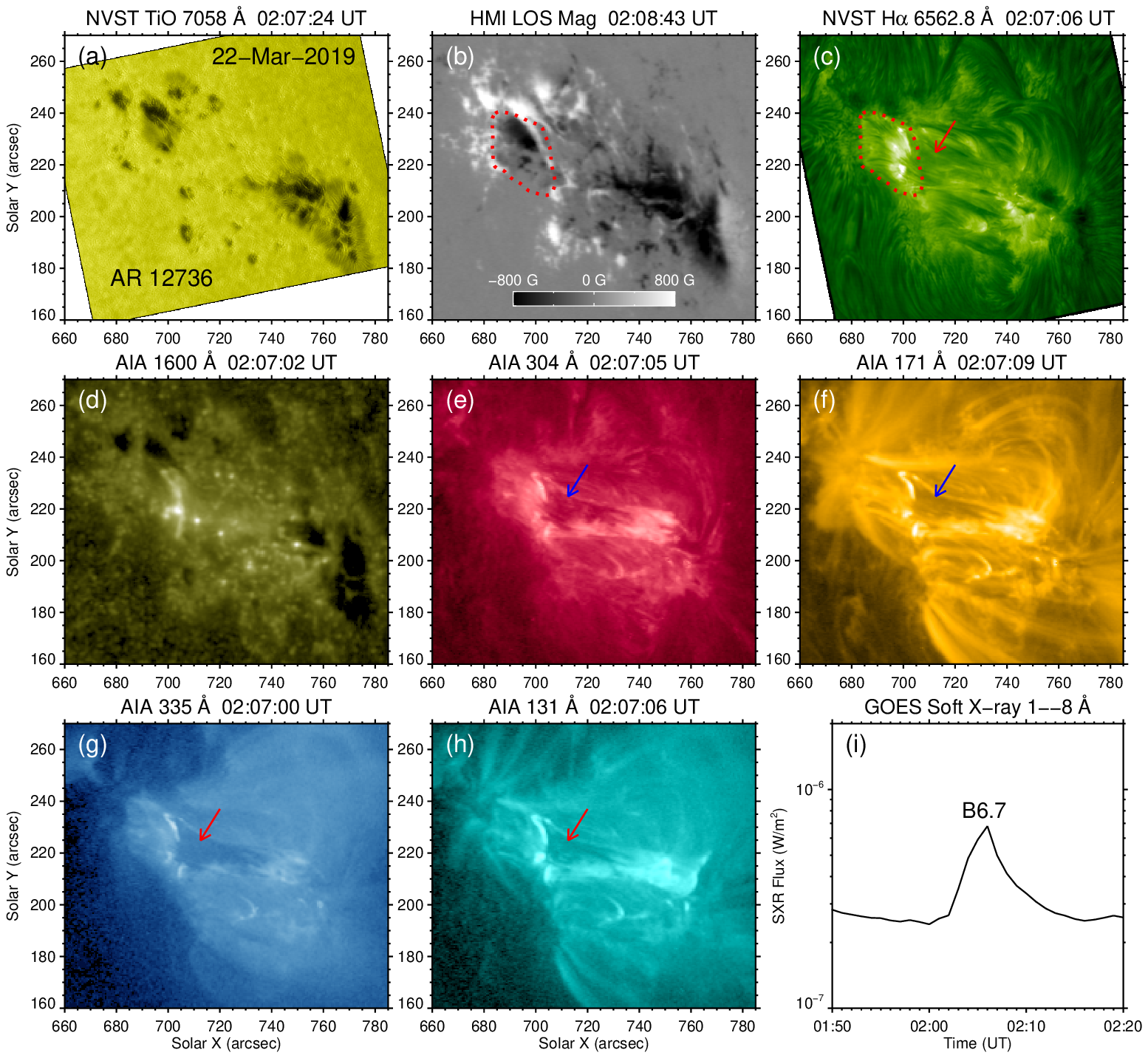} \caption{Overview of AR 12736 during the B6.7 flare on 2019 March 22. Panels (a)-(b): NVST TiO image and HMI LOS magnetogram showing the sunspots and the corresponding magnetic fields, respectively. Panel (c): NVST H$\alpha$ 6562.8 {\AA} image showing the fan-spine structure outlined by the chromospheric material. Panels (d)-(h): AIA 1600 {\AA}, 304 {\AA}, 171 {\AA}, 335 {\AA}, and 131 {\AA} images displaying the appearance of the fan-spine structure at different temperatures. Panel (i): \emph{GOES} soft-X-ray flux in 1--8 {\AA}. The dotted curves in panels (b) and (c) indicate the location of the footpoints of the fan. The arrows in the H$\alpha$ and AIA images denote the dark jet-like structure superposed on the bright spine loops. \protect\\(An animation (Movie1.mp4) of this figure is available.) \label{fig_overview}}
\end{figure*}

AR 12736 was located at about (720{\arcsec}, 220{\arcsec}) on 2019 March 22 (Figure \ref{fig_overview}(a)). Among the trailing sunspots, there were a cluster of negative patches surrounded by positive magnetic fields (delineated by the dotted curve in panel (b)). Within this AR, a B6.7 flare started at 02:00 UT and peaked at 02:06 UT (see panel (i)). After the flare maximum time, the soft X-ray flux decreased gradually until around 02:16 UT. In the H$\alpha$ image, there are several arcades of fibrils tracing partly a fan-spine geometry (panel (c), see also Movie1), which can be shown more clearly, later on, with other wavelengths and the magnetic extrapolation. The footpoints of the fan are indicated by the dotted curve, and the outer spine is denoted by the arrow. In the 1600 {\AA} image (panel (d)), the pronounced brightening coincides with the null point, where the magnetic reconnection took place. In the 304 {\AA}, 171 {\AA}, 335 {\AA}, and 131 {\AA} images (panels (e)-(h)), there is a bright loop bundle connecting the null point and the right remote negative field, tracing the spine field line. Superposed on the bright loop bundle, there is a dark jet-like structure (denoted by the arrows) in different temperature lines.

In order to study the fan-spine magnetic topology, the three-dimensional (3D) coronal magnetic fields are reconstructed using the NLFFF modeling based on the \emph{Hinode}/SP vector magnetogram (Figure \ref{fig_Q}). The photospheric vector magnetic fields (shown in panel (a)) observed by the SP are disambiguated and deprojected. The green rectangle covers the parasitic negative patches surrounded by the opposite polarity. There are two groups of negative patches located nearby the northwestern positive fields (outlined by two ellipses in panel (b)). Around the polarity inversion lines (PILs), the negative and positive fields are highly sheared, as revealed by the arrows in panel (a). Above the PILs, two flux ropes, i.e., flux rope 1 (FR1) and flux rope 2 (FR2) shown in panel (c), are identified in the extrapolated 3D magnetic fields. For FR1, only the field lines with $\mathcal{T}_w > 1$ are presented. For FR2, no field line is twisted more than one turn. Here, we plot the field lines with $\mathcal{T}_w > 0.5$. Therefore, FR2 is actually a group of highly sheared field lines. In the corresponding \emph{Q} map (panel (d)) at $z=0$ (i.e., the photosphere), the high-\emph{Q} ($\thickapprox 10^{4}$) regions (denoted by the arrows) are the intersections of the quasi-separatrix layers (QSL; D{\'e}moulin et al. 1996, 1997) with the photosphere. QSL1 and QSL2 correspond to the PILs, along which FR1 and FR2 are located (panel (e)). The circular high-\emph{Q} region QSL3 embracing QSL1 and QSL2 corresponds to the footpoints of the fan dome within which the two flux ropes are embedded.

Employing the Poincar{\'e} index and Newton-Raphson methods (Guo et al. 2017), we indeed find a single magnetic null point in the NLFFF model. The null point is situated at about 4.0 Mm height above the southwestern side of the parasitic patches ($x = 701{\arcsec}, y = 216{\arcsec}$), as marked by the red dot symbol in Figure \ref{fig_3D}. The top view and side view of the fan-spine skeleton together with two flux ropes are shown in panels (a) and (b) (see also Movie2), respectively. The footpoints of the fan are located in the circular positive fields, and the inclined spine extends to the western negative magnetic fields. Additionally, in panels (c)-(d), some magnetic loops (blue curves) beneath the fan and a set of large-scale field lines (yellow curves) on the external side of the fan surface are shown. The yellow field lines are organized in a sheet-like structure both near the fan and also on the right side, near the spine. This organization is normal as the spine is embedded in a QSL which is a thin volume (Pontin et al. 2016), so the yellow field lines on the right side trace the border of this QSL. Considering the AR location, the LOS (marked by the white arrow) deviates from the radial direction about 50$\degr$, as indicated by the white arrow in panel (d). Viewing along the LOS direction, the materials which are ejected from the reconnection site and move along the fan surface and the outer spine will result in the blueshift and redshift (indicated by the blue and red arrows), respectively.

When examining the NVST H$\alpha$ images, we find there were two filaments, Filament 1 (F1) and Filament 2 (F2) shown in Figure \ref{fig_filaments}(a1)), corresponding to the two extrapolated flux ropes (FR1 and FR2 shown in Figures \ref{fig_Q} and \ref{fig_3D}). In the AIA 171 {\AA} image (Figure \ref{fig_filaments}(b1)), F1 can be also identified, while there was no obvious counterpart for the weaker FR2. The initial brightening (the contour curve) in the AIA 1600 {\AA} image (panel (c1)) was co-spatial with the projected position of the null point (marked by the red ``+" symbol). Then F2 erupted while F1 kept stable (see panel (a2)). The erupting F2 was observed as a dark structure moving toward the west in the H$\alpha$ image (panel (a2)), while it appeared as a bright elongated structure both in the AIA 171 {\AA} (panel (b2)) and 1600 {\AA} (panel (c2)) images.

Since the ejected filament material moved toward the west, we focus on the remote end of the spine (see Figure \ref{fig_remote}). The west end of the outer spine was located in the negative field, as marked by the green symbol in the HMI magnetogram (panel (a)). Before the arrival of the erupted material, there was no brightening in the AIA 1600 {\AA} image (panel (b1)). Around the flare peak time, the erupted material began to hit the chromosphere above the footpoint of the spine, and a faint brightening can be identified, as outlined by the red circle in panel (b2). About one more minute later, the brightness therein was significantly enhanced (see panel (b3)).

Due to the eruption of F2, a jet-like structure (see Figure \ref{fig_rotation}(a)) was formed with filament material moving toward the west. During this process, the jet-like structure rotated, as revealed by the helical motion of dark features. Along slits ``A--B" and ``C--D" marked in panel (a), two time-distance diagrams are derived from the AIA 304 {\AA} images and plotted in panels (b) and (c), respectively. The red dotted curves delineate the trajectories of the dark features with time. The motion patterns of the fine-scale features indicate that the jet-like structure rotated clockwise. This behavior is consistent with the twist sign of FR2, which is the evidence of twist transfer. When FR2 reconnected with the nearby large-scale loops, the twist was transferred from the flux rope to the large-scale system, appearing as the untwisting motion. This scenario is consistent with the events reported by Li et al. (2015) and Xue et al. (2016). Moreover, we can find that the begin times of the violent brightening and obvious rotation at slit ``C--D" were about one minute later than those at slit ``A--B." It means that the twist was propagated along the spine to the remote region.

To illustrate the physical process of this event, we give a schematic drawing (see Figure \ref{fig_cartoon}) based on the observational data and the extrapolated magnetic fields. Initially, there exists a fan-spine magnetic topology (see the green field lines) with a flux rope (the pink twisted lines) within the fan dome. The overlying arcades (represented by the left blue line) of the flux rope and the western anti-directed external loops (the right blue lines) approach each other and reconnect at the null point (marked by the red star symbol). Then a longer bent field line (i.e., the higher yellow field line) and a shorter field line (i.e., the lower yellow line) are formed and consequently move upward and downward, respectively. This is a breakout-type reconnection. As the constraint from the overlying arcades is removed, the flux rope then rises to the null point and reconnects with the western external loops. Due to the magnetic reconnection, material flows move from the reconnection site toward the east/west (represented by the blue/red arrows), resulting in the blueshift/redshift signals, respectively, if seen along the LOS. Since the flux rope is also involved into the reconnection, the twist is transferred to the newly formed large-scale system and propagated to the west along the spine. When the ejected material reaches the remote end of the spine, the lower atmosphere is hit and brightens.

When the flare just started (see Figure \ref{fig_Ha}(a1)), there was a faint brightening nearby the projected position of the null point (marked by ``+"), and the emission at the footpoints (marked by the dotted curve) slightly increased. Around the peak time, the null point had significantly brightened (panel (a2)). As shown in Movie3, some dark material was expelled from the null point and moved bi-directionally to the east and west. At 02:12:29 UT, the fan-spine configuration was partly outlined by dark material (panel (a3)). We derive H$\alpha$ Dopplergrams (panels (b1)-(b3)) from the blue-wing intensity ($I_{b}$) and red-wing intensity ($I_{r}$) with the equation $D=(I_{b} - I_{r})/(I_{b} + I_{r})$, where the negative (positive) value of \emph{D} represents the blueshift (redshift), indicating the material motion toward (away from) the observer (Langangen et al. 2008). At 02:00:53 UT, a pair of blueshift/redshift patches (denoted by the arrows in panel (b1)) existed on the left/right of the reconnection site with the mean velocity of --25/21 km s$^{-1}$. At the flare peak, the reconnection region exhibited a blueshift of about --38 km s$^{-1}$ (panel (b2)). About 6 min later, on the two sides of the reconnection site, there were two large patches with blueshift and redshift of --48 km s$^{-1}$ and 26 km s$^{-1}$, respectively. In the H$\alpha$ velocity movie (Movie3), there exist red/blue extended regions in the northern/southern part of the ejected plasma (towards the right side). The deduced sign of twist is consistent with the twist sign of the pre-eruptive flux rope (see Figure 2). Along curve ``A--B" marked in panel (a1), a time-distance diagram (see panel (c)) is derived from the H$\alpha$ 6562.8 {\AA} images. At the early stage, the dark material moved toward ``A" and ``B" with the projected velocities of about 26 km s$^{-1}$ and 55 km s$^{-1}$, respectively. After the flare peak, the average velocities of the material outflows to ``A" and ``B" are about 37 km s$^{-1}$ and 60 km s$^{-1}$, respectively.

The FOV of the SJI 1330 {\AA} images covered most of the fan-spine structure and the \emph{IRIS} slit crossed the reconnection site (see Figures \ref{fig_absorption}(a)-(b)). In the H$\alpha$ image (panel (a)) at 02:02:12 UT, there was dark chromospheric material located at the null point (marked by ``+"). While in the 1330 {\AA} image, there was only conspicuous brightening at the same position (see panel (b) and Movie4). In the spectral image (panel (c)), the intensity of Si~{\sc{iv}} 1393.76 {\AA} line is obviously enhanced at the null point where reconnection occurred. The corresponding line profile is displayed with the black curve in panel (d). The single Gaussian fitting (blue curve) reveals a blueshift of $-$103 km s$^{-1}$. Additionally, there is an enhancement in the red-wing, as denoted by the red arrow. We average the line profiles within the nearby relatively quiet region, and the resultant profile is shown as the reference profile (the green curve). Compared with the reference profile, the reconnection region is characterized by greatly enhanced and broadened Si~{\sc{iv}} profile. It is prominent that several deep absorption lines (Fe~{\sc{ii}} 1392.15 {\AA}, S~{\sc{i}} 1392.59 {\AA}, Fe~{\sc{ii}} 1392.82 {\AA}, Ni~{\sc{ii}} 1393.33 {\AA}, and Si~{\sc{iv}} 1393.76 {\AA}) are superposed on the Si~{\sc{iv}} profile. All the five absorption lines have the same blueshift of $-$0.10 {\AA}, i.e., $-$22 km s$^{-1}$, compared with their theoretical positions.

At the flare peak time, the reconnection region violently brightened in 1330 {\AA} and crossed by the \emph{IRIS} slit (Figure \ref{fig_spec_peak}(a)). On the right side, there was a bright jet-like structure extending to the west. Along ``A--B" crossing the reconnection region (marked in panel (a)), a time-distance diagram (see panel (b)) is derived. During the flare, there were several intermittent bright outflows (outlined by the dotted lines) from the reconnection site toward ``B" with the mean velocity of 67 km s$^{-1}$. In the spectral image (panel (c)), some positions (marked by ``+") are obviously blueshifted or redshifted. As denoted by the arrow, there exists a prominent C-shaped dark feature corresponding to Ni~{\sc{ii}} 1393.33 {\AA} line. Particularly, at position ``P1", Ni~{\sc{ii}} line has a large blueshift. The Si~{\sc{iv}} line profile at ``P1" is plotted as the black curve in panel (d1). The single Gaussian fitting (blue curve) reveals a blueshift of $-$129 km s$^{-1}$. The arrow denotes the Ni~{\sc{ii}} 1393.33 {\AA} deep absorption superposed on the Si~{\sc{iv}} profile. Compared with the position of the steady Ni~{\sc{ii}} line, the observed absorption line has a blueshift of $-$38 km s$^{-1}$. At ``P2," the Si~{\sc{iv}} emission (see panel (d2)) is about ten times stronger than that at ``P1" and ``P3." The double Gaussian fitting shows a blueshift component of $-$71 km s$^{-1}$ and a redshift component of 128 km s$^{-1}$. The Ni~{\sc{ii}} absorption line is shallow and non-shifted. The Si~{\sc{iv}} profile at ``P3" is also double Gaussian fitted well (see panel (d3)). It reveals the existence of a blueshift of $-$127 km s$^{-1}$.

\section{Conclusions and Discussion}

With the simultaneous observations from the NVST, \emph{SDO}, \emph{IRIS}, and \emph{Hinode}, we investigated an inclined fan-spine structure responsible for a B6.7 flare. When the flare started, the fan-spine brightened in the UV and EUV channels. In the H$\alpha$ line, it was partly filled and outlined by dark outflows. We extrapolated the coronal magnetic field and confirmed the existence of the fan-spine magnetic topology. When magnetic reconnection occurred at the null point, some dark material was ejected from the reconnection site and moved along the fan dome and spine lines, resulting in the blueshift and redshift in the H$\alpha$ Dopplergrams, respectively. At the reconnection region, the Si~{\sc{iv}} line profile was enhanced both in the blue-wing and red-wing, revealing bi-directional outflows. Superposed on the Si~{\sc{iv}} profile, there were several blueshifted deep absorption lines.

\begin{figure*}
\centering
{\subfigure{\includegraphics[bb=107 291 493 536,clip,angle=0,width=0.82\textwidth]{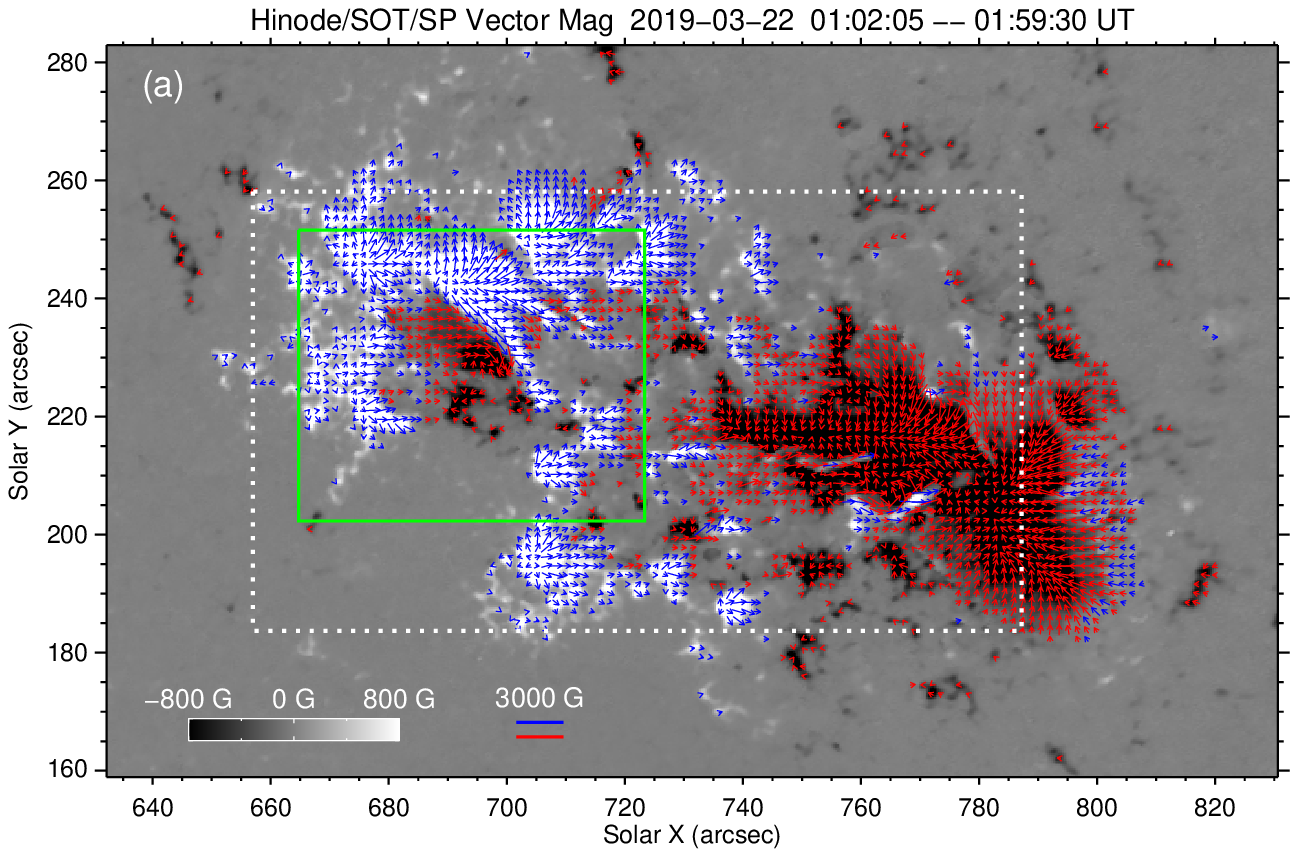}} \quad
\subfigure{\includegraphics[bb=107 284 493 564,clip,width=0.82\textwidth]{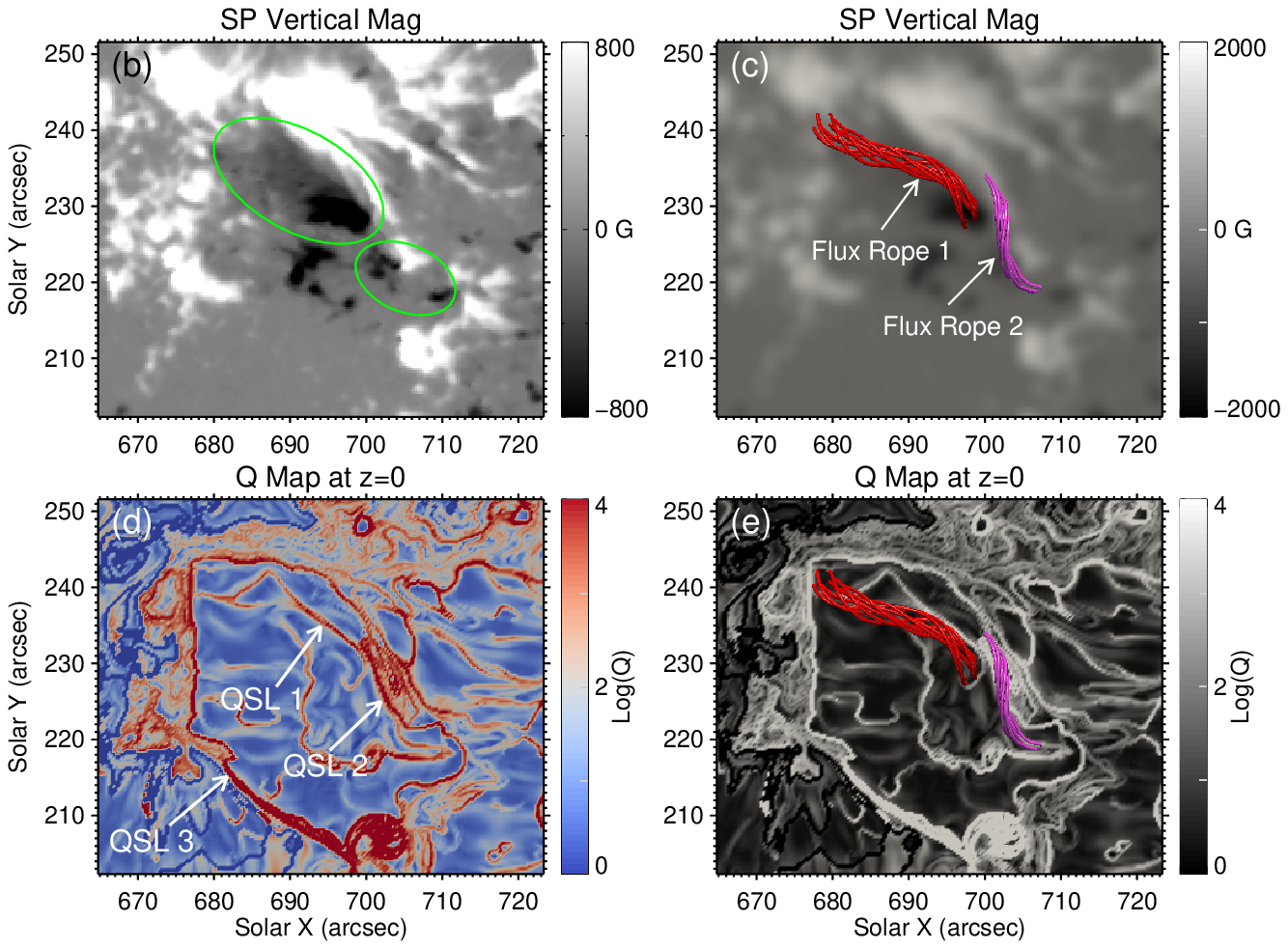}}}
\caption{Panel (a): photospheric vector magnetic fields observed by \emph{Hinode}/SP. Panels (b)-(c): sub-FOV of SP magnetogram together with two NLFFF extrapolated magnetic flux ropes. Panels (d)-(e): corresponding squashing factor \emph{Q} map together with two flux ropes (red and pink ones). The orientation and length of each blue/red arrow in panel (a) indicate the direction and strength of the horizontal field at a given position. The green and white rectangles outline the FOVs of panels (b)-(e) and Figure \ref{fig_3D}, respectively. The ellipses in panel (b) outline two groups of negative magnetic patches responsible for two flux ropes. \label{fig_Q}}
\end{figure*}

For a flux rope, there is a critical twist, above which it will be unstable. This kind of instability is termed kink instability, and the typical threshold value is about 1.75 turns, i.e., 3.5$\pi$ (Hood \& Priest 1979; T{\"o}r{\"o}k \& Kliem 2003, 2005; Kumar et al. 2012). In this event, the maximum twist number of the flux rope (i.e., the pink flux rope in Figure \ref{fig_Q}) involved into the flare was only $\sim 2\pi$. Therefore, the eruption here should not be triggered by the kink instability. The images in Figure \ref{fig_filaments} show that the initial brightening was spatially coincided with the null point, implying the magnetic reconnection first took place at the null point. We propose that the breakout-type reconnection led to the eruption. The typical magnetic configuration of the magnetic breakout model is a multi-polar field (Antiochos 1998; Antiochos et al. 1999), which has been widely reported in observations and well studied in simulations (Lynch et al. 2004; DeVore \& Antiochos 2008; Jiang et al. 2013; Chen et al. 2016). The sketch in Figure \ref{fig_cartoon} based on the breakout reconnection model can explain the observed phenomena well.

\begin{figure*}
\centering
\includegraphics
[bb=44 262 546 566,clip,width=1.0\textwidth]{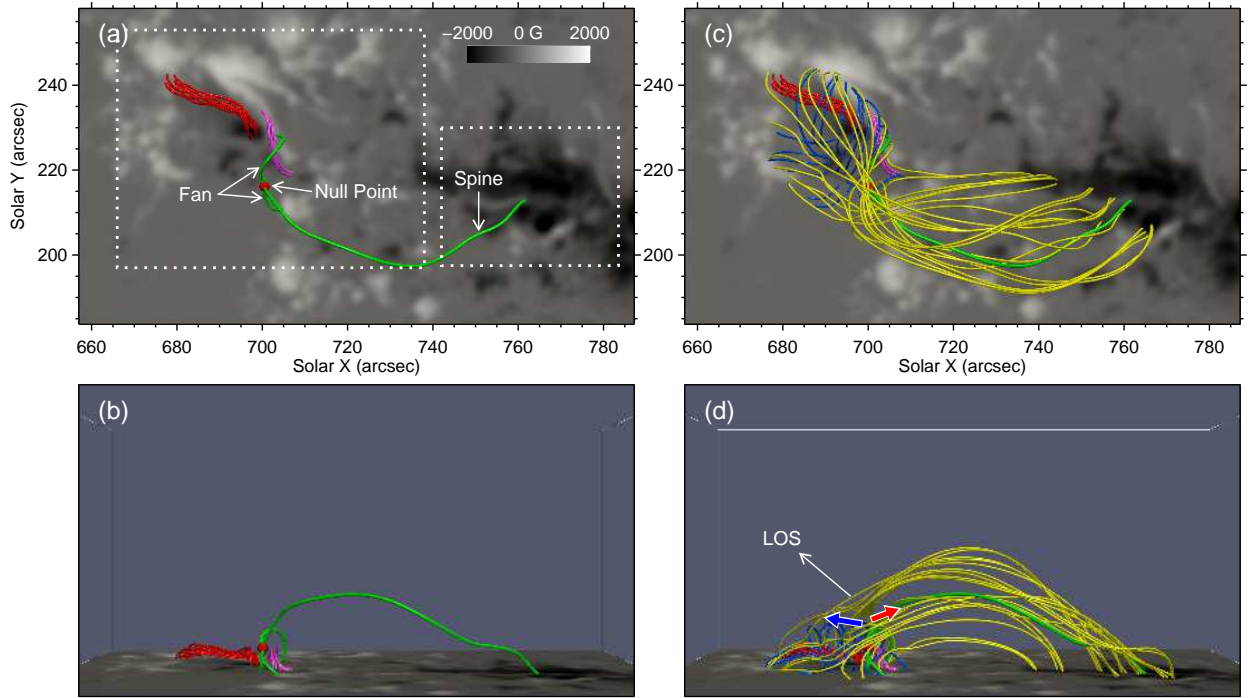} \caption{Panels (a)-(b): top view and side view of the skeleton of the fan-spine (green curves) and the null point (red dot symbol) together with two flux ropes, respectively. Panels (c)-(d): top view and side view of the reconstructed coronal structures, respectively. The large and small rectangles in panel (a) outline the FOVs of Figures \ref{fig_filaments} and \ref{fig_remote}, respectively. The white arrow in panel (d) denotes the LOS direction, and the blue and red arrows indicate the material flows resulting in the blueshift and redshift, respectively. \protect\\(An animation (Movie2.mp4) of this figure is available.) \label{fig_3D}}
\end{figure*}

\begin{figure*}
\centering
\includegraphics
[bb=73 262 514 574,clip,width=1.0\textwidth]{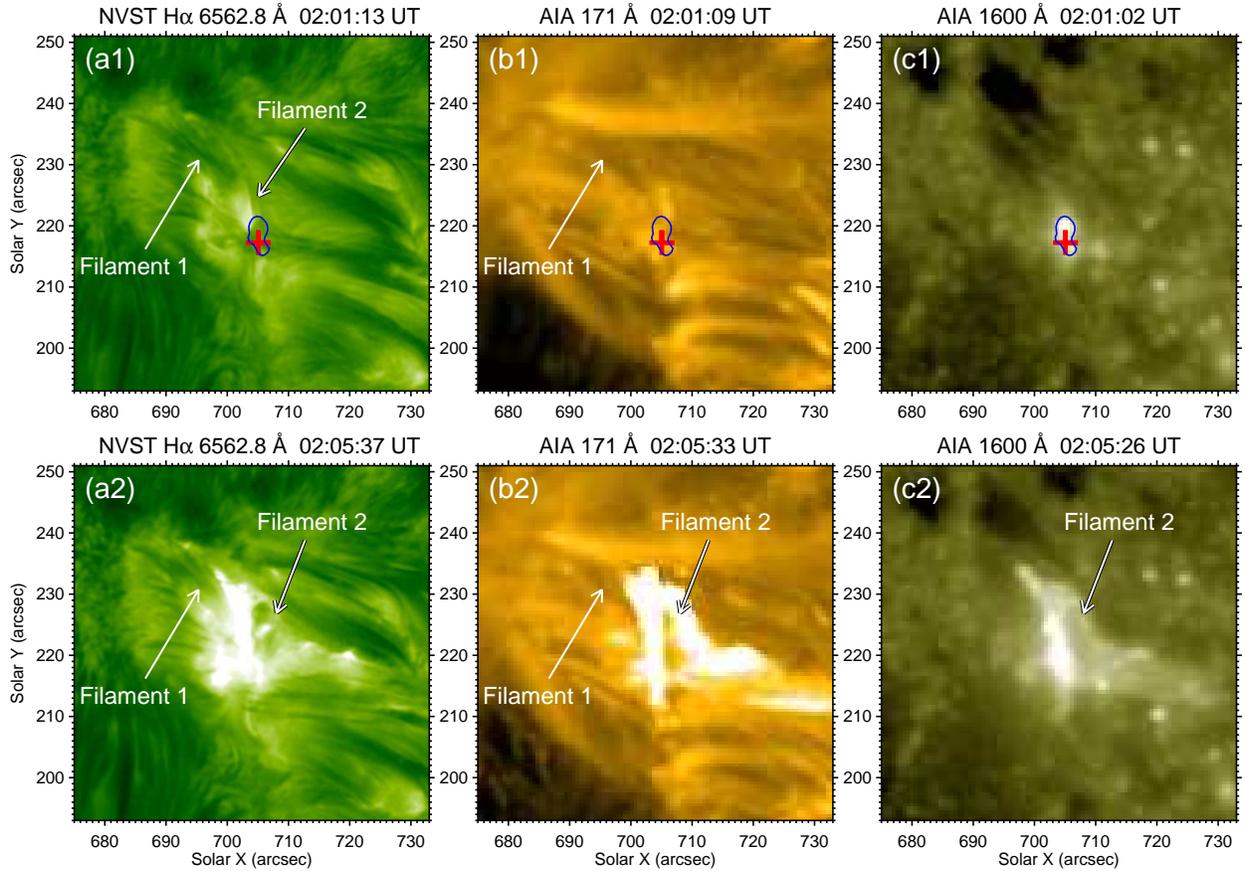} \caption{NVST H$\alpha$ 6562.8 {\AA} (left column), AIA 171 {\AA} (middle column), and 1600 {\AA} (right column) images before (upper panels) and during (lower panels) the filament eruption. In the upper panels, the red ``+" symbols mark the projected position of the null point, and the blue curves are the contour of the significant brightening in the AIA 1600 {\AA}. \label{fig_filaments}}
\end{figure*}

\begin{figure*}
\centering
\includegraphics
[bb=180 294 429 560,clip,width=0.5\textwidth]{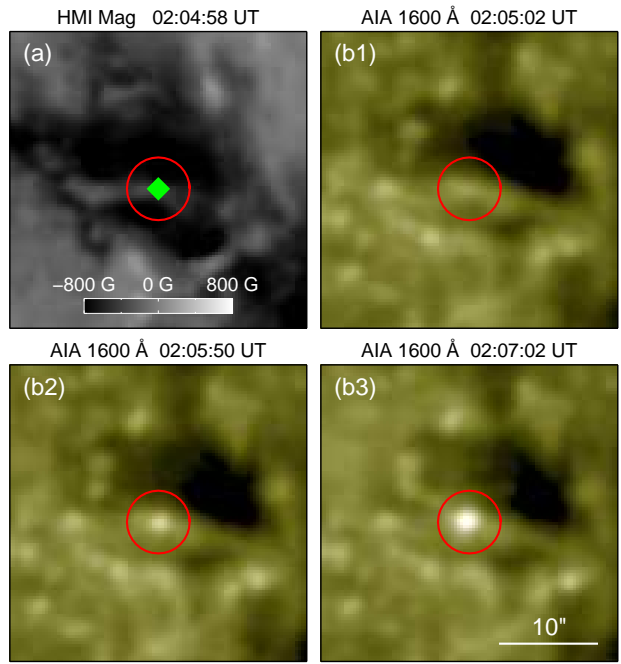} \caption{HMI magnetogram (panel (a)) and the corresponding AIA 1600 {\AA} images (panels (b1)-(b3)) at the remote end of the spine. The green symbol marks the position of the footpoint of the outer spine, and the red circles outline the brightening region in the AIA 1600 {\AA}. \label{fig_remote}}
\end{figure*}

\begin{figure*}
\centering
\includegraphics
[bb=184 185 406 651,clip,width=0.45\textwidth]{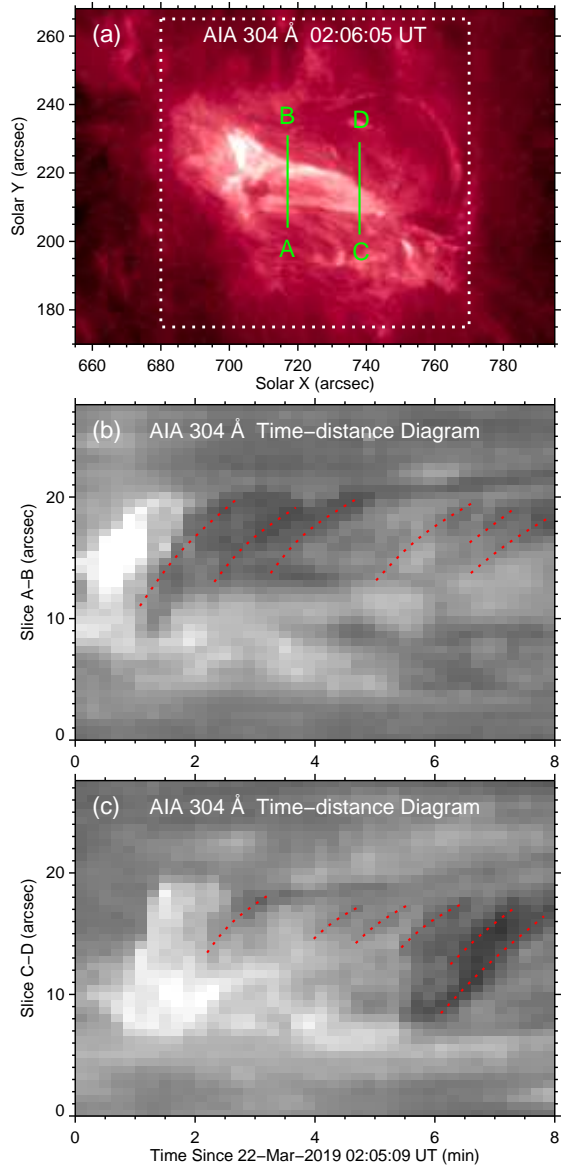} \caption{Panel (a): AIA 304 {\AA} image around the flare peak time showing the jet-like structure resulting from the filament eruption. Panels (b)-(c): time-distance diagrams derived from the AIA 304 {\AA} images along ``A--B" and ``C--D" (marked in panel (a)), respectively. The dotted rectangle in panel (a) outlines the FOV of Figure \ref{fig_Ha}, and the dotted curves in panels (b)-(c) delineate the trajectories of the dark features with time. \label{fig_rotation}}
\end{figure*}

\begin{figure*}
\centering
\includegraphics
[clip,width=0.9\textwidth]{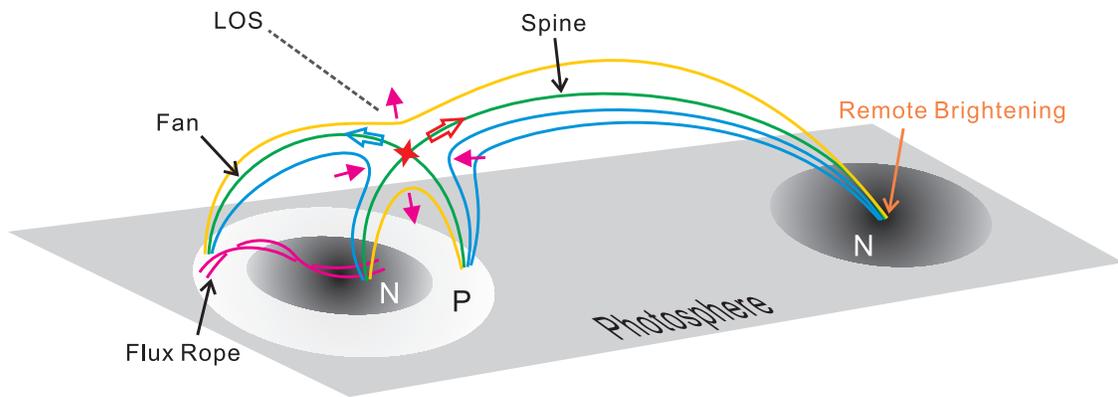} \caption{Sketch of the magnetic structures illustrating the physical process of the observed event. The white and black patches represent the positive (P) and negative (N) magnetic fields, respectively, which are connected by the fan-spine topology (green field lines). At the null point (marked by the red star symbol), two sets of blue field lines undergo breakout-type reconnection, and two sets of anti-directed new loops (yellow field lines) are formed. The approach and separation directions are denoted by the pink arrows. The accelerated materials move outward from the reconnection site, resulting in the blueshift and redshift (represented by the blue and red bold arrows) relative to the LOS direction, respectively. \label{fig_cartoon}}
\end{figure*}

\begin{figure*}
\centering
\includegraphics
[bb=102 198 484 636,clip,width=0.85\textwidth]{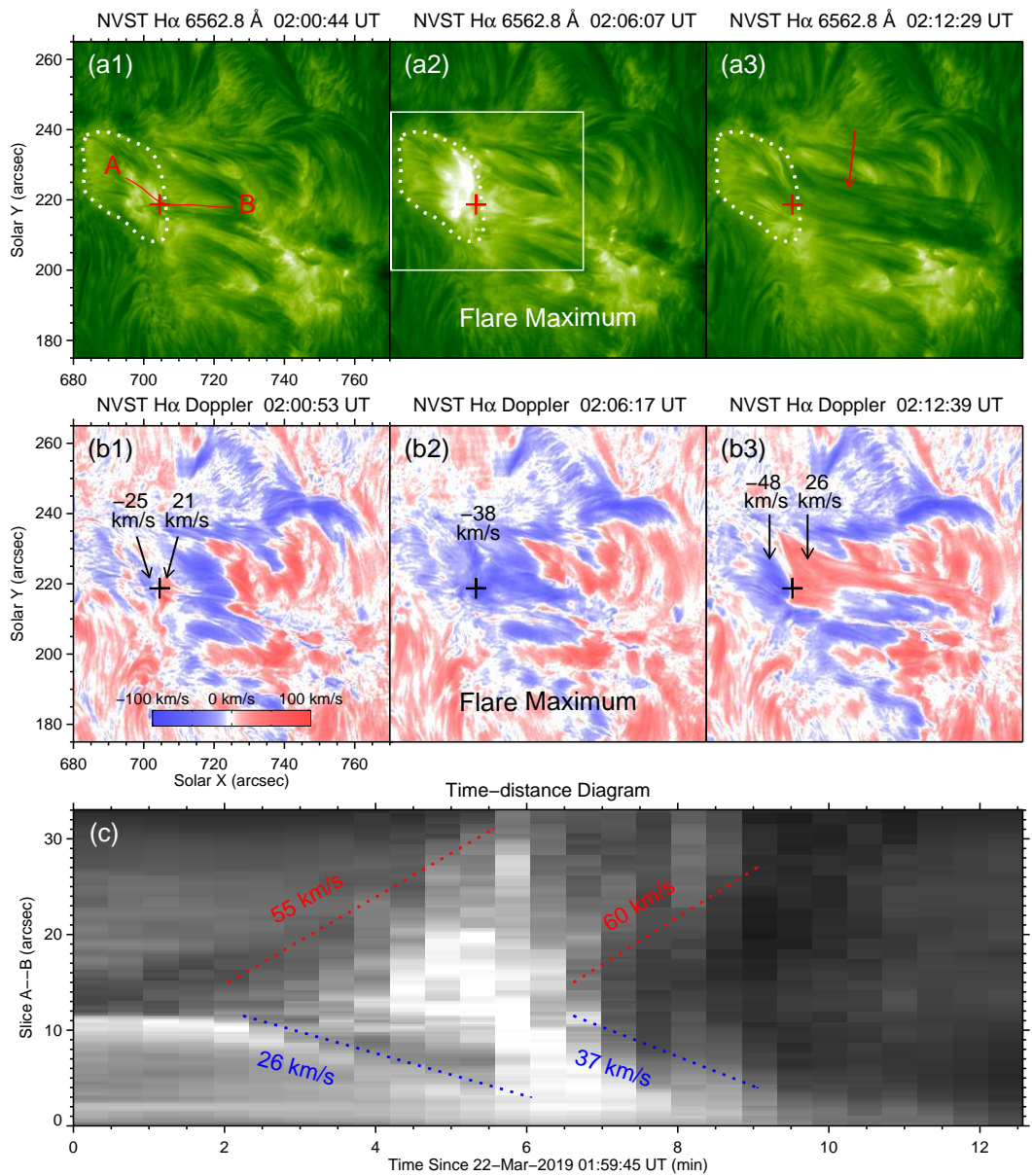} \caption{Sequence of H$\alpha$ 6562.8 {\AA} images (panels (a1)-(a3)) and the corresponding H$\alpha$ Dopplergrams (panels (b1)-(b3)). The ``+" symbols mark the projected position of the null point where magnetic reconnection occurred, the dotted curves outline the footpoints of the fan dome, and the arrow in panel (a3) indicates the spine loops. The arrows in panels (b1) and (b3) denote the patches with conspicuous blueshift and redshift. The rectangle in panel (a2) outlines the FOV of Figure \ref{fig_absorption}. Panel (c): time-distance diagram derived from the H$\alpha$ images along ``A--B" marked in panel (a1). The dotted lines delineate the dark outflows from the reconnection site. \protect\\(An animation (Movie3.mp4) of this figure is available.) \label{fig_Ha}}
\end{figure*}

During the flare, the most violent brightenings in the NVST, AIA, and SJI images spatially coincide with the null point, indicating the occurrence of the null-point reconnection (Priest \& Pontin 2009) in the fan-spine topology. In the H$\alpha$ Dopplergrams shown in Figures \ref{fig_Ha}(b1) and (b3), the material flows to the east along the fan surface and to the west along the outer spine correspond to the blueshift and redshift signals, respectively. This is consistent with the projection of the fan-spine configuration relative to the LOS (see Figure \ref{fig_3D}(d)). Combing the LOS velocity (Figures \ref{fig_Ha}(b1)-(b3)) and the projected velocity in the sky plane (Figure \ref{fig_Ha}(c)), we can obtain the total velocity of the material flows. Before the flare peak, the dark material moved to the east along the fan with the total velocity of 36 km s$^{-1}$ and to the west along the spine with the total velocity of 59 km s$^{-1}$. After the flare peak, these two velocities are 61 km s$^{-1}$ and 65 km s$^{-1}$, respectively. When the flux rope (i.e., FR2 shown in Figures \ref{fig_Q} and \ref{fig_3D}, also that in Figure \ref{fig_cartoon}) within the fan dome reconnects with the outer ambient field lines, a set of longer bent field lines (represented by the higher yellow field line in Figure \ref{fig_cartoon}) will be formed and consequently lifted upward, as indicated by the upward pink arrow. This can be used to explain why the reconnection region around the peak time corresponds to a blueshift patch in the H$\alpha$ Dopplergram (see panel (b2)).

\begin{figure*}
\centering
\includegraphics
[bb=102 255 488 580,clip,width=0.85\textwidth]{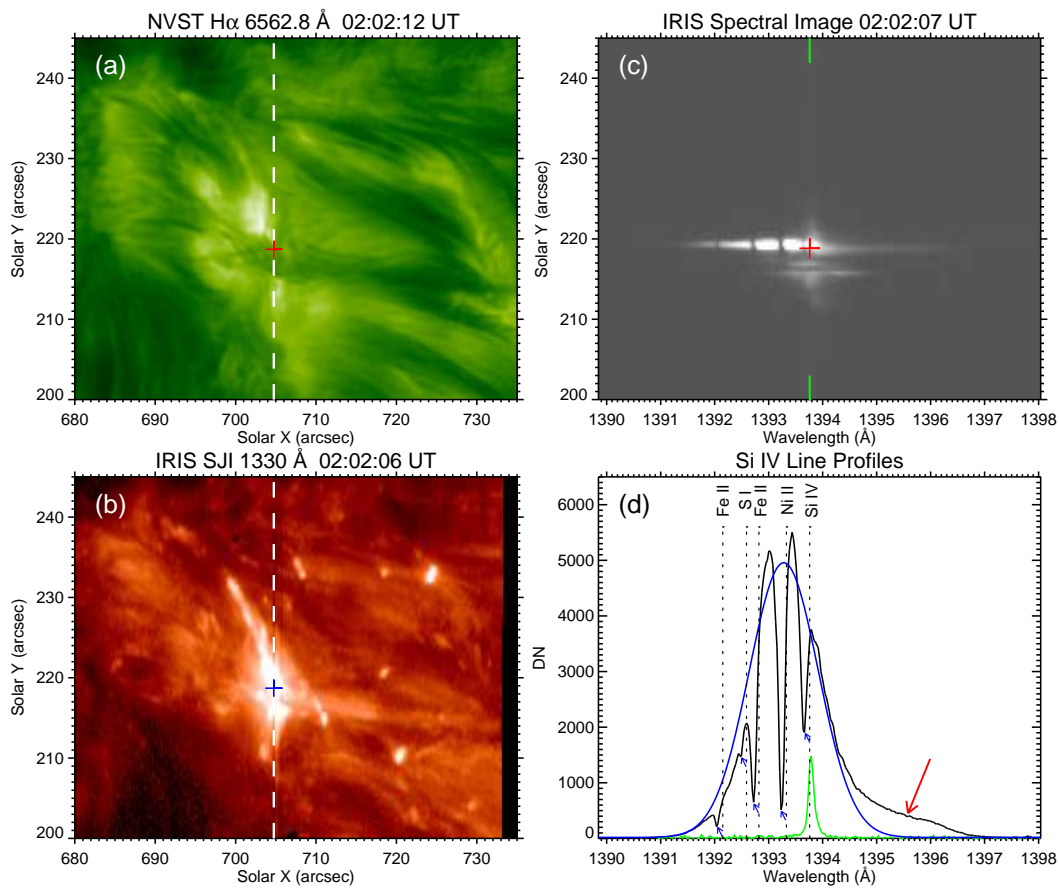} \caption{Panels (a) and (b): NVST H$\alpha$ 6562.8 {\AA} and \emph{IRIS} SJI 1330 {\AA} images at 02:02 UT. The ``+" symbols mark the null point. Panel (c): \emph{IRIS} spectral image of Si~{\sc{iv}} along the slit marked by the dashed lines in panels (a) and (b). The two green vertical bars mark the position of 1393.76 {\AA}. Panel (d): observed Si~{\sc{iv}} line profile (black curve) at the reconnection region and the corresponding single Gaussian fitting (blue curve). The red arrow denotes an enhancement in the red-wing of the Si~{\sc{iv}} profile, and the blue arrows indicate the blueshifted absorption lines. The reference spectrum (green curve) is the averaged profile in the relatively quiet region, the values of which have been multiplied by 50. \protect\\(An animation (Movie4.mp4) of this figure is available.) \label{fig_absorption}}
\end{figure*}

\begin{figure*}
\centering
\includegraphics
[bb=102 174 487 661,clip,width=0.85\textwidth]{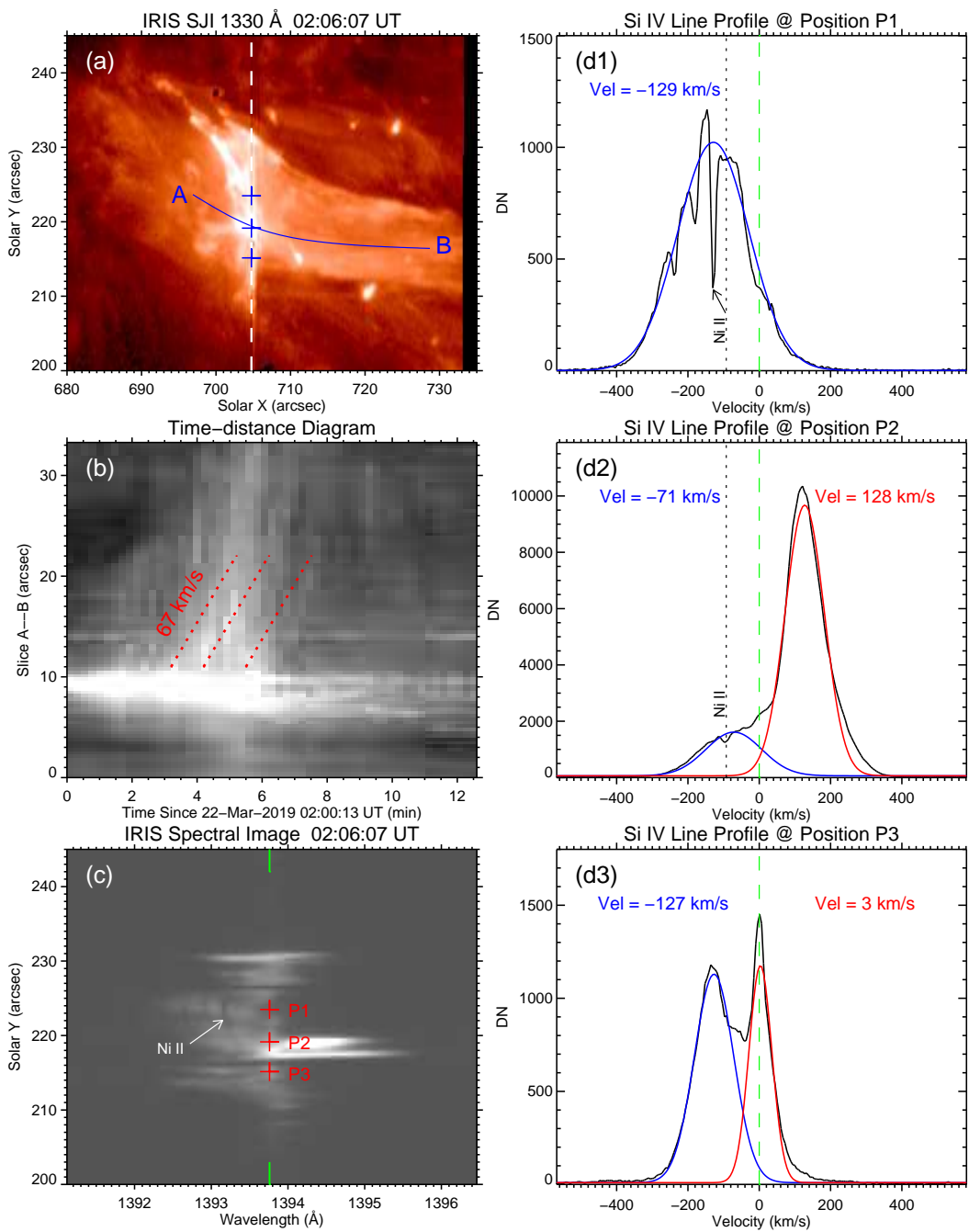} \caption{Panel (a): \emph{IRIS} SJI 1330 {\AA} image at the peak time of the flare. Panel (b): time-distance diagram along ``A--B" marked in panel (a). The dotted lines outline the intermittent outflow from the reconnection site toward ``B." Panel (c): \emph{IRIS} spectral image of Si~{\sc{iv}} along the slit marked by the dashed lines in panel (a). Panels (d1)-(d3): Si~{\sc{iv}} 1393.76 {\AA} line profiles (black curves) at three positions marked by ``+" symbols in panels (a) and (c). The blue curve in panel (d1) is the single Gaussian fitting, and the blue and red curves in panels (d2) and (d3) are the double Gaussian fitting. \label{fig_spec_peak}}
\end{figure*}

At the null point, both the blue-wing and red-wing of the \emph{IRIS} Si~{\sc{iv}} line profile are enhanced during the flare (Figures \ref{fig_absorption}(d) and \ref{fig_spec_peak}(d2)), implying the existence of bi-directional outflows due to magnetic reconnection. As shown in Figure \ref{fig_spec_peak}(b), the trajectories of the bright outflows along the spine loops can be well determined. Considering their projected velocity in the sky plane is 67 km s$^{-1}$ and the value of the redshift is 128 km s$^{-1}$, thus the total velocity is 144 km s$^{-1}$. At two positions (``P1" and ``P3") nearby the reconnection site (``P2"), the Si~{\sc{iv}} line profiles are blueshifted (see Figures \ref{fig_spec_peak}(d1) and (d3)). Since these two positions are located on the fan dome instead of on the null point, the blueshifts at ``P1" and ``P3" are speculated to result from the material flows passing ``P1" and ``P3" to the east along the fan or may be due to upward lifting of the newly formed large-scale loops.

One prominent phenomenon shown in Figure \ref{fig_absorption}(d) is the superposition of several deep absorption lines on the Si~{\sc{iv}} line profile. The presence of these absorption lines indicates that there exists cooler material above the hot reconnection region (Tian et al. 2018). When examining the H$\alpha$ and SJI images, we find that there is a dark feature in the H$\alpha$ image (Figure \ref{fig_absorption}(a)) and a brightening in the 1330 {\AA} image (Figure \ref{fig_absorption}(b)) at the reconnection site. Thus the 1330 {\AA} brightening is inferred to be located below the H$\alpha$ chromospheric material. The observed profile shows that all the five absorption lines are blueshifted with the same value of $-$0.10 {\AA}, which should be due to the movement of the cooler material toward the observer. Therefore we deem that the dark material observed in the H$\alpha$ image was moving toward us with the velocity of 22 km s$^{-1}$, which is used to convert the digital numbers in the H$\alpha$ Dopplergrams to the H$\alpha$ Doppler velocity (see Figures \ref{fig_Ha}(b1)-(b3)). Moreover, some absorptions (e.g., Ni~{\sc{ii}} line) are quite deep, which also depends on the amount of cooler material overlying the null point. These results imply that such spectral profiles with the superposition of absorption lines can be used as a tool to diagnose the properties of cooler material above the reconnection site.

\acknowledgments {We are grateful to the referee for the constructive comments and valuable suggestions. We thank Prof. Hui Tian and Dr. Yuandeng Shen for helpful discussion. This work is supported by the National Key R\&D Program of China (2019YFA0405000), the National Natural Science Foundations of China (11673035, 11773079, 11873091, 11790304, 11533008, 11790300), the B-type Strategic Priority Program of the Chinese Academy of Sciences (XDB41000000), the Key Programs of the Chinese Academy of Sciences (QYZDJ-SSW-SLH050), and the Youth Innovation Promotion Association of CAS. Z.Z. and Y.G. are supported by NSFC (11773016, 11733003, 11533005, 11961131002), and the China Scholarship Council 201906190107. The data are used courtesy of NVST, \emph{IRIS}, \emph{Hinode}, \emph{SDO}, and \emph{GOES} science teams. \\ }

{}

\clearpage

\clearpage

\clearpage

\clearpage

\clearpage

\clearpage

\clearpage

\clearpage

\clearpage

\clearpage

\clearpage

\end{document}